\documentclass{elsart}
\usepackage{graphicx}
\usepackage{amssymb}

\begin{document}
\journal{Solid State Communications}
\begin{frontmatter}

\title{Aging of a nanostructured Zn$_{50}$Se$_{50}$ alloy produced by mechanical alloying}
\author{K. D. Machado\corauthref{cor1}},
\ead{kleber@fisica.ufsc.br}
\corauth[cor1]{Corresponding author.}
\author{J. C. de Lima},
\author{C. E. M. de Campos},
\author{T. A. Grandi},
\author{A. A. M. Gasperini}
\address{Depto de F\'{\i}sica, Universidade Federal de Santa Catarina, Trindade, Cx. P. 476, 
88040-900, Florian\'opolis, Santa Catarina, Brazil}

\begin{abstract}

The aging at room temperature of a nanocrystalline equiatomic ZnSe alloy produced by mechanical 
alloying was investigated using X-ray diffraction (XRD) and differential scanning calorimetry (DSC) 
techniques. The measured XRD patterns showed the presence of the peaks corresponding to the 
crystalline trigonal selenium ({\em c}-Se) phase. It is observed that the ZnSe phase is stable 
with aging. The arising of the {\em c}-Se phase is attributed to the migration of Se atoms located 
at the interfacial component of the as-milled ZnSe nanostructure with aging. 
\end{abstract}

\end{frontmatter}

\section{Introduction}
\label{}

Due to their physical and chemical characteristics, selenium-based alloys are very important from  
technological and scientific points of view, mainly those alloys containing germanium or zinc 
because of their very interesting optical properties. For example, zinc selenide (ZnSe) is used in 
infrared lenses field. More recently, researches on semiconductor quantum dots have shown that it can 
also be used in short-wavelength visible-light laser devices \cite{Leppert}. 
According to them, the 
use of ZnSe in this field needs that the alloy is synthesized in nanometer form, with crystallite 
sizes smaller than the ZnSe exciton Bohr diameter ($\approx 90$ \AA). For this purpose, techniques 
such as arrested nucleation in glasses, precipitation from sol-gel solutions and entrapment in 
porous sites inside zeolite cavities have been employed. These techniques, beyond expensive, are not 
able to full control the size of the nanoparticles obtained.

Mechanical alloying (MA) \cite{Mec} has been used for almost two decades to produce many unique materials. 
These include, for instance, nanostructured alloys, amorphous compounds and unstable and metastable 
phases \cite{Leppert,Lima1,Weeber,Froes,Yavari,carlos1,carlos2,kleber}. 
This method has several intrinsic advantages, like low temperature processing, easy 
control of composition, relatively inexpensive equipment, and the possibility of scaling up. 
Although the MA technique is relatively simple, the physical mechanisms involved are not yet fully 
understood. In order to make use of this technique in industrial applications, a better understanding 
of these physical mechanisms is desirable.

From a structural point of view, nanostructured materials can be regarded as being made of two 
components, one crystalline, with dimensions of the order of some nanometers, that preserves the 
structure of bulk crystal, and one interfacial, composed by defect centers. This component has 
caused controversy in the literature. Gleiter \cite{Gleiter2} has described it based on a gaseous 
model while 
other authors \cite{Stern2} disagree. The number of atoms in both components is similar 
\cite{Grandi2}. Due 
to this fact, 
nanostructured material properties are strongly dependent of the interfacial component. From a 
technological point of view, manipulation of the interfacial component makes it possible to 
design materials with desired physical properties for specific applications 
\cite{Grandi2,Grandi}.

The Zn-Se phase diagram \cite{Anderko} shows only an equiatomic ZnSe phase. In a 
recent paper \cite{Lima2} de Lima {\em et al}. reported structural results obtained from the 
application of MA to Zn$_{1-x}$Se$_x$ ($x = 0.20$, 0.30, 0.40, 0.50, 0.70 at. \%) mixtures. The 
measured X-ray diffraction (XRD) patterns for the mixtures with nominal compositions different of 
the equiatomic one showed the peaks associated to the ZnSe phase and those ones associated to the 
majority component. In the case of Se-rich mixtures, Se excess was found in amorphous state. More 
recently, in other paper \cite{Joao3}, de Lima {\em et al}. reported results on the influence of 
aging in the structural properties of the amorphous selenium ({\em a}-Se) produced by ball 
milling (BM). They have observed that the atomic structure of {\em a}-Se produced by BM, which is 
formed by Se$_n$ chains, has transformed partially to another known {\em a}-Se structure formed by 
Se$_8$ rings found in the amorphous state obtained by rapid quenching or vapor deposition 
methods. Based on this interesting observation, an aged nanostructured ZnSe sample produced by 
MA some years ago was re-examined by using XRD and DSC techniques. To our knowledge, it is the first 
time that such study is reported.

\section{Experimental procedure}

A nanostructured equiatomic ZnSe sample (hereafter called ZnSe-97) was prepared by MA in 1997 
following the procedure described in Ref. \cite{Lima2}. The X-ray diffraction (XRD) pattern measured 
after milling was indexed to the cubic zinc selenide phase (isostructural with the structure of ZnS 
compound), with lattice parameter $a = 5.6478$ \AA. After four years being kept in a dry desiccator 
a new XRD pattern of the same sample (hereafter called aged-ZnSe-97) was measured in order to 
verify possible structural changes. The comparison between the XRD patterns of the as-milled and 
aged samples shows significant differences that will be discussed later. To ensure reproductibility 
of the ZnSe-97, a new equiatomic ZnSe sample was prepared by MA. Its XRD pattern was identical to 
that recorded for the as-milled ZnSe-97 sample.

Small quantities of the aged-ZnSe-97 sample were heat treated at 196$^\circ$C and 350$^\circ$C 
in quartz capsules containing argon. Only the sample heated at 350$^\circ$C was quenched. The other 
was air-cooled. During the heat treatment at 350$^\circ$C, formation of water drops at the inner 
walls of the capsule was observed, indicating that the samples are hydrophilic. All heated samples 
were analyzed using a Rigaku powder diffractometer, Miniflex model, with Cu K$_\alpha$ radiation 
($\lambda = 1.5418$ \AA) and a TA 2010 Differential Scanning Calorimetry (DSC) cell with a heating 
rate of 10$^\circ$C/min in a flowing argon atmosphere.

\section{Results and discussion}

\begin{figure}
\begin{center}
\includegraphics{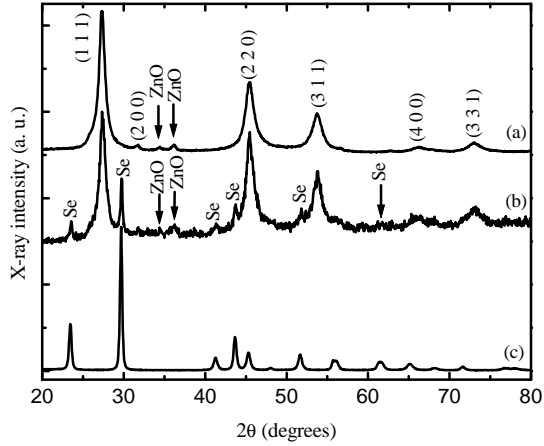}
\caption{\label{fig1} XRD patterns for (a) ZnSe-97, (b) aged-ZnSe-97 and (c) crystalline Se phase.}
\end{center}
\end{figure}

The measured XRD patterns for ZnSe-97, aged-ZnSe-97 and crystalline selenium in its trigonal form 
({\em c}-Se) are shown in Figs. \ref{fig1}.a, \ref{fig1}.b and \ref{fig1}.c, respectively. In 
a previous study \cite{Lima2} we showed that the XRD pattern seen in Fig. \ref{fig1}.a corresponds to 
the nanostructured ZnSe compound. A comparison between Figs. \ref{fig1}.a and \ref{fig1}.b shows that 
this phase is also seen in the aged-ZnSe-97 sample. This indicates that the ZnSe phase is as stable 
with aging as expected. In addition to the peaks associated with the ZnSe phase, Fig. \ref{fig1}.b 
displays several other peaks that were indexed to the {\em c}-Se phase and that are absent in 
Fig. \ref{fig1}.a. The remaining low intensity peaks seen in Figs. \ref{fig1}.a and \ref{fig1}.b 
were indexed to a contaminant ZnO phase according to the JCPDS card No. 1-1136 \cite{jcpds}. 
Their low intensities when compared to those associated with the ZnSe phase suggested that this 
contaminant should be found in a small quantity, and a Rietveld analysis \cite{Rietveld} 
(discussed below) confirmed this assumption. 

\begin{figure}
\begin{center}
\includegraphics{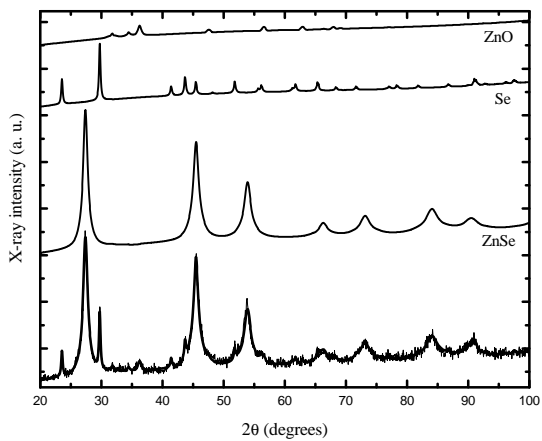}
\caption{\label{fig2} XRD pattern of the aged-ZnSe-97 (bottom) and its Rietveld fitting (thick line) 
using ZnSe, {\em c}-Se and ZnO phases.}
\end{center}
\end{figure}

The ZnSe-97 and aged-ZnSe-97 patterns shown in Figs. \ref{fig1}.a and \ref{fig1}.b were simulated 
using the Rietveld procedure \cite{Rietveld} considering ZnSe, {\em c}-Se and ZnO phases. The 
results obtained show that the lattice parameter of the ZnSe phase changes with aging within 
0.1\% and their values are listed in Table \ref{tab1}. Besides that, about 93\% of the crystalline 
phases are given by the ZnSe phase and 7\% by the ZnO phase in the as-milled sample, whereas in the 
aged sample the ZnSe, ZnO and {\em c}-Se phases are responsible for about 
73\%, 16\% and 11\% of the phases, 
respectively (see Fig. \ref{fig2}). By using the full width of half maximum (FWHM) obtained from the 
simulated patterns and the Scherrer formula \cite{Klug} the average crystallite sizes of the 
phases were estimated. The results for the ZnSe phase are found in Table \ref{tab1}. Its value for 
the ZnO phase is 129 \AA\ in the as-milled and 135 \AA\ in the aged samples. In the aged sample, 
the {\em c}-Se phase has an average crystallite size of 335 \AA, indicating that all phases are 
found in the nanometric form. As mentioned above, nanostructured alloys are described by two 
components, one crystalline and other interfacial. Thus, the segregation of the {\em c}-Se particles 
in the aged sample could have had its origin in the migration of Se atoms located at the interfacial 
component and, if this is true, it is a remarkable fact since the {\em c}-Se segregation process 
has occurred at room temperature. According to Gleiter \cite{Gleiter2}, the interfacial component 
is formed by a high density of interfaces ($ \approx 10^{19}$ cm$^{-3}$), which gives raise to a 
high density of short diffusion paths. As a consequence of the presence of an interfacial component 
in the nanostructured material, it is expected an increase in the diffusion coefficient in 
comparison with single crystals and polycrystals with the same chemical composition. For instance, 
according to Birringer {\em et al}. \cite{Birringer}, the self-diffusivity in nanocrystalline copper 
(80 \AA) at 353 K is $10^{-18}$ m$^2$/s while in the lattice its value is $10^{-34}$ m$^2$/s. 
Partial amorphous structural transformation with aging at room temperature had also been observed in 
amorphous selenium produced by BM \cite{Joao3}.

\begin{table}
\caption{\label{tab1} Lattice parameters and average crystallite sizes of ZnSe phase prepared by MA.}
\begin{tabular}{ccc}
Sample & Lattice parameter (\AA) & Average crystallite size (\AA)\\
ZnSe-97 & 5.6478  & 96 \\
aged-ZnSe-97 & 5.6408 & 92 \\
aged-ZnSe-97 (treat. at 196$^\circ$C) & 5.6443 & 107\\
aged-ZnSe-97 (treat. at 350$^\circ$C) & 5.6578 & 189\\
JCPDS card. No. 37-1463& 5.6688 & ---
\end{tabular}
\end{table}

\begin{figure}[h]
\begin{center}
\includegraphics{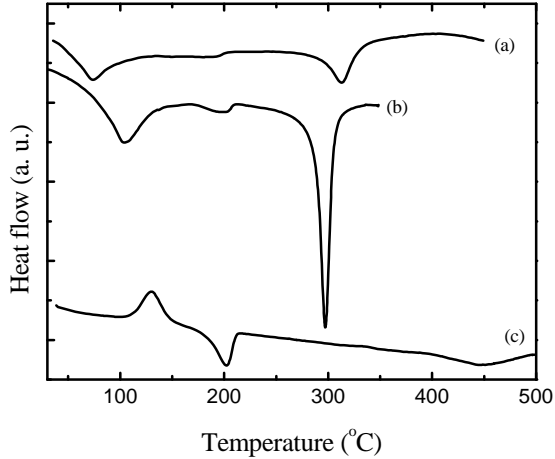}
\caption{\label{fig3} DSC spectra for (a) ZnSe-97, (b) aged-ZnSe-97 and (c) aged-ZnSe-97 after 
heat treatment at 350$^\circ$C.}
\end{center}
\end{figure}

To investigate the presence of residual non-reacted Se in the ZnSe-97 sample, which could have not 
been detected by the XRD measurement, a DSC measurement was performed and it is shown in 
Fig. \ref{fig3}.a. It displays three endothermic peaks located at 75$^\circ$C, 200$^\circ$C and 
310$^\circ$C. These peaks, except that at 200$^\circ$C, were associated with water, as will be 
discussed later. The peak at 200$^\circ$C is associated with the melting of a very small quantity of 
{\em c}-Se phase. These {\em c}-Se particles could have acted as ``seeds" for the growing of the 
{\em c}-Se phase identified in the aged-ZnSe-97 (see Fig. \ref{fig1}.b). Fig. \ref{fig3}.b, which 
corresponds to the DSC curve of the aged-ZnSe-97 sample, shows three endothermic peaks located at 
100$^\circ$C, 200$^\circ$C and 300$^\circ$C. As it happened to the ZnSe-97 sample, the peaks at 
100$^\circ$C and 300$^\circ$C were associated with water. The peak at 200$^\circ$C in this figure 
is higher than that in Fig. \ref{fig3}.a suggesting the growing of the {\em c}-Se phase, in 
agreement with the XRD results.

To determine the origin of the peaks located around 75--100$^\circ$C and 300$^\circ$C in the DSC 
measurements of ZnSe-97 and aged-ZnSe-97 (see Figs. \ref{fig3}.a and \ref{fig3}.b) we performed a 
heat treatment in a small quantity of the aged-ZnSe-97 sample at 350$^\circ$C followed by a 
quenching in iced water. After it was made we noted the formation of water drops at the inner walls 
of the capsule, indicating that the sample was hydrophilic. The DSC measurement of the quenched 
sample shown in Fig. \ref{fig3}.c confirmed this assumption since in this curve the peaks located 
around 75--100$^\circ$C and 300$^\circ$C were not seen anymore. These peaks can be explained by the 
release and vaporization of adsorbed water at the surface and inside of the crystallites, 
respectively. The increase in the intensity of the peaks in the aged-ZnSe-97 DSC spectrum is due to 
the quantity of water adsorbed during aging. The exothermic peak about 130$^\circ$C and the 
endothermic one at 200$^\circ$C (see Fig. \ref{fig3}.c) are associated with the amorphous-crystalline 
Se transformation and with the melting of the {\em c}-Se formed, respectively.

\begin{figure}[h]
\begin{center}
\includegraphics{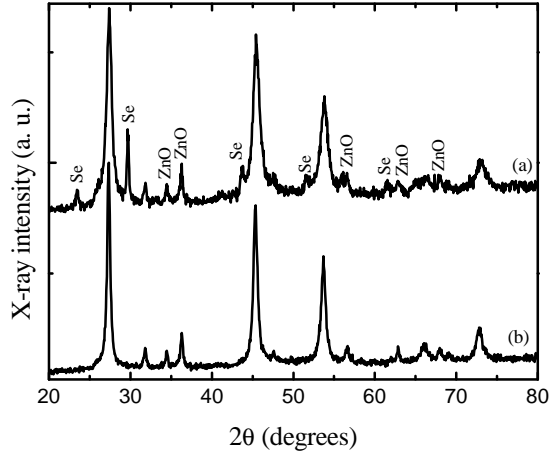}
\caption{\label{fig4} XRD patterns for aged-ZnSe-97 heat treated at (a) 196$^\circ$C and (b) 
350$^\circ$C.}
\end{center}
\end{figure}

In order to investigate the influence of the temperature in the growing of the {\em c}-Se phase 
observed in the aged-ZnSe-97 sample, another small quantity of the aged-ZnSe-97 sample was heat 
treated at 196$^\circ$C. At this temperature water was not observed at the inner walls of the 
capsule, indicating that almost all water observed in the sample quenched at 350$^\circ$C was 
released from inside the crystallites. The measured XRD pattern of the sample heat treated at 
196$^\circ$C is shown in Fig. \ref{fig4}.a. Its comparison with that of the aged-ZnSe-97 sample 
displayed in Fig. \ref{fig1}.b reveals that this heat treatment has only caused a slight improvement 
in the crystallinity of the ZnSe phase, as it can be seen by comparing the lattice parameters and 
average crystallite size values given in Table \ref{tab1}. The Rietveld analysis indicates that there 
is no change in the phase quantities with the heat treatment, therefore suggesting that the 
alloy had already reached its thermodynamic stability before the heat treatment. The crystallinity 
of the ZnSe phase is significantly improved only at 350$^\circ$C, as indicated by its XRD pattern (see Fig. 
\ref{fig4}.b) and by its lattice parameter given in Table \ref{tab1}. As expected, the XRD pattern 
of the quenched sample does not show the peaks associated with {\em c}-Se anymore, in agreement with 
the DSC measurement (see Fig. \ref{fig3}.c), which shows that the Se particles are found in an 
amorphous phase. In addition, Rietveld analysis indicates that the ZnSe and ZnO phases correspond 
to about 
84\% and 16\% of the crystalline phases found in this sample, respectively. These data indicate 
an increase in the quantity of the ZnO phase as compared to that in the ZnSe-97 sample. The 
average crystallite size of the ZnO phase also increases, reaching 268 \AA.

\section{Conclusion}

The influence of aging in a nanocrystalline equiatomic ZnSe alloy prepared by MA was investigated. 
Based on the results found, we conclude that:

\begin{enumerate}
\item The nanocrystalline equiatomic ZnSe alloy produced by MA is as stable as those prepared 
by using other methods (melting, for instance), as indicated by the XRD patterns seen in Figs. 
\ref{fig1} and \ref{fig4}. 

\item Due to the presence of a small quantity of {\em c}-Se phase in the as-milled sample and to the 
properties of the interfacial region already described in the text, the room temperature 
energy ($\approx 0.025$ eV) is sufficient to promote migration of Se atoms located at this component. 
This fact causes growing of the {\em c}-Se phase, as it could be seen in the aged-ZnSe-97 XRD 
pattern (Fig. \ref{fig1}.b).

\item After four years the growing of the {\em c}-Se phase has finished, and the crystalline 
phases in the sample have reached their thermodynamic stability, as indicated by the XRD 
measurements and Rietveld analyses of the aged- and heat-treated at 196$^\circ$C ZnSe-97 
samples (Figs. \ref{fig1}.b and \ref{fig4}.a).

\item The samples are hydrophilic, as indicated by the DSC measurements shown in Fig. \ref{fig3}.

\end{enumerate}

\ack

We thank to the Brazilian agencies CNPq, CAPES and FINEP for financial support.


\end{document}